\documentclass[a4paper,fleqn,usenatbib,useAMS]{mnras}
\usepackage{graphicx,dblfloatfix}   
\usepackage{amsmath}    
\usepackage{amssymb}    
\usepackage{multicol}        
\usepackage{bm}     
\usepackage{pdflscape}  
\usepackage{xcolor}
\usepackage{ae,aecompl}
\usepackage[english]{babel}
\usepackage{array, booktabs, multirow}

\catcode`_=\active
\newcommand_[1]{\ensuremath{\sb{\mathrm{#1}}}}
\catcode`^=\active
\newcommand^[1]{\ensuremath{\sp{\mathrm{#1}}}}

\newcommand{\jlike}{ Jovian }
\newcommand{\einr}{ R_{\rm{E}}}

\title[WFIRST observations of Microlensing]{Detection of Exoplanet as a Binary Source of Microlensing Events in WFIRST Survey}
\author[Bagheri et al.]{ Fatemeh Bagheri$~^{1}$\thanks{E-mail: bagheri.fateme@gmail.com },
                                  Sedighe Sajadian$~^{2}$\thanks{E-mail: s.sajadian@cc.iut.ac.ir},
                                  Sohrab Rahvar$~^{1}$\thanks{E-mail: rahvar@sharif.edu }
                                           \footnotemark[1] \\
$^{1}$Department of Physics, Sharif University of Technology, P.O.
Box 11155-9161, Tehran, Iran\\
$^{2}$Department of Physics, Isfahan University of Technology, Isfahan 84156-83111, Iran}

\date{Accepted XXX. Received YYY; in original form ZZZ}

\pubyear{2019}

\begin{document}
\label{firstpage}
\pagerange{\pageref{firstpage}--\pageref{lastpage}}
\maketitle

\begin{abstract}
    We investigate the possibility of exoplanet detection orbiting source stars in microlensing events through WFIRST observations. We perform a Monto Carlo simulation on the detection rate of exoplanets via microlensing, assuming that each source star has at least one exoplanet. The exoplanet can reflect part of the light from the parent star or emit internal thermal radiation. In this new detection channel, we use microlensing as an amplifier to magnify the reflection light from the planet. In the literature, this mode of detecting exoplanets has been investigated much less than the usual mode in which the exoplanets are considered as one companion in binary lens events.
    Assuming $72$ days of observation per season with the cadence of $15$ minutes, we find the probability of rocky planet detection with this method to be virtually zero. However, there is non-zero probability, for the detection of Jovian planets. We estimate the detection rates of the exoplanets by this method, using WFIRST observation to be $0.012\%$ in single lens events and $0.9\%$ in the binary lens events.
\end{abstract}


\section{Introduction}
    \label{sec:intro} \label{sec:intro}
The study and detection of exoplanets, as well as the investigation of their structures and potential for habitability, has become of paramount importance and interest to the astronomical community in the past decade. Various detection methods such as transit, radial velocity, astrometry, microlensing and direct imaging have already explored thousands of exoplanets \citep{knutson2014friends, mandel2002analytic,seager2003unique,traub2010direct,guyon2005exoplanet,Gaudi2012,rahvar2015gravitational}. Despite the technological advances in exoplanet detection based on space-based telescopes, microlensing is still the only feasible approach for detecting exoplanets deep into the Milky Way Galaxy, beyond the immediate neighborhood of the Earth as well as planets beyond the snow line of the parent stars.\\
The standard approach in gravitational microlensing method for exoplanet detection is based on the assumption that the lens star has an exoplanet, thus forming a binary lens. The caustic crossing of this binary system with the source star makes a deviation from the simple microlensing light curve. There is another method in microlensing that is proposed by \citet{2000ApJ...538L.133G} and \citet{sajadian2010illuminating} where the planet orbiting around the source star can reflection light from its parent star and makes a binary source.\\
The signature of the planet as the second source can be detected either (i) by the caustic crossing in the binary lensing system (which is also proposed for detecting Extra Terrestrial Intelligent (ETI) with the radio-wavelength \citep{rahvar2016gravitational} follow-up microlensing observation ) or (ii) in the single high magnification events. So far, this mode of detecting exoplanets via microlensing events has not been explored as in the standard mode, however there have been papers on the detection of exoplanets in microlensing events in which exoplanets are considered as sources \citep{2005ApJ...628..478S, 2003ApJ...586..527G, 2001A&A...380..292L, 2001MNRAS.325..305A, 2000ApJ...539L..63L}. In this work, we investigate the prediction of detecting exoplanets that bounded with the source stars.
Toward this goal, in this paper, we investigate the possibility and the rate of the detection of exoplanets orbiting the source star in single-lens microlensing events as well as in a binary-lens systems using WFIRST space-based telescopes.\\
To simulate the light curves for single/binary-lens configurations, we assume the WFIRST's microlensing survey would monitor in the direction of $b = -1.5$ and $l = 0.5$ in the Galactic coordinate with the cadence of $15$-minutes in $72$-day observation per season.
We assume the finite source effect for the source star but no limb darkening \citep{gould1996finite}. The plan of the paper is as follows: In section \ref{sec:methods}, we discuss the methodology for simulation of microlensing events,  potentially detectable by WFIRST. The results will be summarized in section \ref{sec:results}. Implications of the results, conclusion, and  prospects of exoplanet detection via high-magnification single-lens and binary-lens microlensing events will be discussed in section \ref{sec:discussion}.

\section{Methodology}
    \label{sec:methods}

\subsection{WFIRST Specification and Observation Strategy}
The Wide Field Infrared Survey Telescope (WFIRST) is a conceptual NASA satellite. Besides addressing important questions in a wide range of topics in astrophysics and cosmology, a primary science objective of WFIRST is to study the demographics of planetary systems and to detect exoplanets via multiple methods, such as direct imaging (based on the coronagraph onboard WFIRST), and microlensing. The $2.4$ m telescope design of WFIRST has a smaller field of view than the $\approx 1 m$ class designs, but can monitor significantly fainter stars at a given photometric precision due to their smaller PSF and larger collecting area, resulting in a similar number of fields being required to reach the same number of stars, despite the difference in field of view. After these two effects cancel, the designs with larger diameter mirrors come out as significantly more capable scientifically due to their improvement in the ability to measure relative lens-source proper motions. Investigation of this possibility is even more promising when knowing that WFIRST $W149$ filter ($0.9-2 \mu m $) will have a significantly lower detection threshold, with the zero-point magnitude of $27.61$.\\
Despite the technology implemented in WFIRST coronagraph, exoplanet detection by direct imaging is almost impossible beyond distances of $50$ parsec. This makes microlensing the only feasible approach to detecting exoplanets deep into the Milky Way Galaxy beyond the immediate neighborhood of the Earth. The microlensing survey of WFIRST would monitor $1.97$ $deg^2$ of the Galactic bulge, with $15$-minute cadence, over six $72$-day per season \citep{2015arXiv150303757S}, potentially detecting thousands of exoplanets via the perturbations that they produce on the microlensing light curves.\\
As mentioned in section (\ref{sec:intro}), the underlying assumption in our simulations is that there is an exoplanet orbiting the source star. Therefore, the observed flux is the sum of fluxes of the source star and its planet in addition to background noise. We assume that the noise is composed of the four components: (i) the intrinsic Poisson fluctuation for the flux received from the microlensing event in one exposure $\sigma_{\star}$, (ii) the intrinsic Poisson fluctuation inside the PSF representing the background sky flux in one exposure $\sigma_{\rm{sky}}$, (iii) the read-out noise $\sigma_{\rm{read}}$, and (iv) the dark noise $\sigma_{\rm{dark}}$. Thus, the total noise $\sigma_{\rm{tot}}$ in the observed flux is given by,
    \begin{eqnarray}\label{eq:noise}
        \sigma_{\rm{tot}}=\sqrt{\sigma_{\star}^{2}~+~\sigma_{\rm{sky}}^{2}~+~\sigma_{\rm{read}}^{2}~+~\sigma_{\rm{dark}}^{2}}~.
    \end{eqnarray}

Then we calculate the interval that the noisy signal has magnitude less than the zero-point magnitude of WFIRST and also more than saturation magnitude which is $\approx 14.8$ \citep{2019ApJS..241....3P}. Since WFIRST is a wide field telescope (with a view $100$ times greater than Hubble's), there is no need that events to be alerted to WFIRST. The exposure time for the filter $W149$ is $46.8 s$ and the average of slew time and settle time is $83.1 s$.

\subsection{Parameters for the source and lens stars}

The simple microlensing effect (a point-like source and a point-like lens with uniform relative motion with respect to the line of sight) has some characteristic features that allow one to distinguish it from any known intrinsic stellar variability. These features are as follows: given the low probability of the source and detector alignment within $\einr$ (Einstein radius), the event should be singular in the history of the source (as well as of the deflector's). The magnification is independent of the color and is a simple function of time, depending on the minimum impact parameter (i.e., $u_{\rm{0}}$), the Einstein crossing time (i.e., $t_{\rm{E}}$, which is the time for crossing the Einstein radius) and the time of maximum magnification (i.e., $t_{\rm{0}}$). As the geometric configuration of the source-deflector system is random, the impact parameters of the events must be uniformly distributed. The so-called simple microlensing description can be complicated in many different ways: for instance, multiple lens and source systems, extended sources, parallax, and Xallarap effects (\citet{moniez2017understanding} and the references therein). These complications are beyond the scope of this study and have no significant effect on the results. 
In this paper, we consider extended source and therefore, define the source star radii where the normalized angular size of the star to the Einstein angle is called $\rho_{\star}$. In the case of binary-lens microlensing events, we need two extra parameters which determine the mass ratio of the mass lenses, $q$, and the separation distance between them, $s$.\\
For mass density in the Milky Way galaxy, we use the Besan\c{c}on model as discussed in \citep{schlegel1998maps, robin2003synthetic, gardner2014n}. In this model, the distribution of the matter in our Galaxy is described by the superposition of the eight thin disk structures with different ages, a thick disk component, and a central (old) bar structure made of two components. We consider the updated model from \citep{gardner2014n} that appears to be specifically adapted to the Galactic plane, and choose the fitted parameters associated with a two ellipsoid bar. This model with using the distribution of stars from the Hipparcos catalog has been successfully used to interpret the microlensing data of EROS collaboration in the direction of the spiral arms \citep{moniez2017understanding}.
For generating binary lenses, the mass ratio between the two lenses is taken from the distribution function  \citep{duquennoy1991multiplicity} as
\begin{eqnarray}
\centering
\xi(q) = \exp{\left[{\frac{-(q-\mu)^2}{2\sigma_{q}^{2}}}\right]}
\end{eqnarray}
where $q = M_{2}/M_{1}$, $\mu = 0.23$ and $\sigma_{q} = 0.42$. We choose the semi-major axis $s$ of the binary orbit of microlenses from the {\"O}pik's law where the distribution function for the primary-secondary distance is proportional to $\rho(s) = dN/ds \propto s^{-1}$ in the range of $s \in [0.6, 30]~\rm{AU}$.\\  
In our simulation, we consider the distance effect of the source stars as well as the reddening of the source stars on the stellar apparent magnitudes. After generating the position and the type of a star, we estimate the extinction due to dust along the line of sight using the 3D extinction map well provided by \citep{Marshal2006}. Then, we use the relations in \citep{nishiyama2006interstellar, nishiyama2008interstellar} and \citep{nishiyama2009interstellar} to transpose the extinction of $I$ and $V$ passbands into the passband of WFIRST satellite. We consider extinction in $H$-band for the filter W149 since according to Figure 1 in \citep{2019ApJS..241....3P} this filter has more overlap in H-band.

\subsection{Parameters for the Exoplanets}
In this part we discuss the parameters of the exoplanets, assuming that all source stars have exoplanets. To assign the characteristic parameters of the exoplanets in our simulations, we divide them into two main categories of the rocky and the \jlike planets. Based on the standard theory of the planet formation, the orbital distance of the rocky planets is within the snow line of its parent star \citep{kennedy2006planet,kennedy2008planet}. The snow line of any star can be written in terms of the snow line of the Sun as,
    \begin{eqnarray}
        \centering
        R_{\rm{SL}, \star} = R_{\rm{SL},\odot}\times
        \frac{M_{\star}}{M_\odot}~,
    \end{eqnarray}
where $M_{\rm{\star}}$ is the mass of the parent star \citep{gould2010frequency}. For the \jlike planets, because they can migrate inward from the large distance \citep{1998Sci...279...69M,2006RPPh...69..119P}, we consider the minimum distance close to the parent star equal to $0.01~\rm{A.U.}$ and its maximum amount $10~\rm{A.U.}$. This range is adapted from the MOA microlensing observations that there is at least one bound planet per star in this range, i.e., $a_p = 0.01-10~\rm{A.U.}$ \citep{sumi2011unbound}. We assume that the distribution of the exoplanet semi major axis obeys Opik law.\\
The exoplanet masses have to be in the range of $0.1- 10^{4} M_{\oplus}$. Thus, if the exoplanet is located at the distance closer than the snow-line of its parent star and also its mass $M_{p}$ is less than $6 M_\oplus$ (equivalently, the planet radii $R_{p}$ is approximately less than $1.6 R_\oplus$), then we assume that it is a rocky planet, otherwise it is considered as a \jlike planet. \citep{marcy2014masses, rogers2015most, lopez2016predictions,lopez2014understanding}.
For the binary lenses in our simulation, out of $10^6$  events, $579559$ simulated events are due to binary lenses with a companion having a lower mass in the range of $13-80$ Jupiter mass which is the mass range for brown dwarfs. Therefore, binary lenses with a brown dwarf companion are dominated. Also, in our simulation, we adapt the relation between the radii of the planet and its mass as suggested in \cite{2017A&A...604A..83B} where above $\sim 120$ Earth mass planets, the radius of planets slightly change with their masses.\\
The configuration that is assumed for the lens, source star, and the planet in this paper implies that the observed light curve of microlensing events is a combination of two individual single-lens light curves, one due to the source star and the other by the planet. Thus, in order to add the two light curves, a new parameter representing the ratio of the planet's flux to the flux of the parent star, $z = F_{\rm{p}}/F_{\star}$ is needed. This parameter is one of the main factors in the detectability of the planet.\\
The flux receiving from a planet contains thermal radiation due to the intrinsic temperature of the planet, as well as the reflection radiation from the parent star. Assuming the thermal emission of the planet to follow a black body radiation \citep{lopez2007thermal}, the planet's temperature can be computed by taking into account the absorption of the parent star's radiation by the planet, which is subsequently reradiated according to the Boltzman's law,
    \begin{eqnarray}
        \centering
        T_{\rm{p}}=T_{\rm{eff}} \sqrt{\frac{R_{\star}}{a_{\rm{p}}}} \bigg[ f\times{(1-A_B)} \bigg]^{1/4} ~,
    \end{eqnarray}
where $A_B$ is the albedo, and $f$ describes the fraction of reradiated energy that is absorbed by the planet, $R_\star$ is the radius of the star and $a_p$ is the distance of the planet from the parent star, $T_{eff}$ is the effective temperature of the star and $T_p$ is the temperature of the planet.
(Harrington et al. 2006; Knutson et al. 2007). $f = 2/3$ represents a hot Jupiter with low advection and significant temperature difference between the day and night on the planet. At the other extreme, $f=1/4$ represents a planet with high redistribution of energy. Because a hot Jupiter could be mostly detected, we choose $2/3$ for all the planets. We assume that the Albedo of the rocky and \jlike planets is that of Mars and Jupiter (i.e., $0.15$ and $0.52$), respectively.\\
Assuming that the planet radiation manners same as a black body, its thermal flux ($F_{\rm{th}}$) at the given frequency $\nu$ is given by the Planck's law as,
\begin{eqnarray}\label{eq:power}
\centering
I(\nu,~T_{\rm{p}})=~\frac{2h~\nu^3}{c^2}~\frac{1}{\exp{(h\nu/kT_{\rm{p}})}-1},
\end{eqnarray}
where $I(\nu, Tp)$ represents the emitted power per unit area of the emitting surface, per unit solid angle, per unit frequency. Integrating \eqref{eq:power} over the energy window of WFIRST detector (W149 filter) and over an half-sphere yields the thermal flux of the planet as seen by WFIRST. The flux of
the planet due to the reflection (i.e., $F_{\rm{ref}}$) of parent star’s light is given by
    \begin{eqnarray}\label{flux}
        F_{\rm{ref}}= A_g~g(\Phi)~F_{\star}(\frac{R_{\rm{p}}}{a_{\rm{p}}})^{2},
    \end{eqnarray}
where $g(\Phi)$ indicates a fraction of the lighted area of the planet in front of the observer, $A_g$ is geometric albedo, $F_{\star}$ is the flux of the parent star, $R_{\rm{p}}$ is the radius of the planet. For geometric albedo, we use the relation $A_g = 2/3 A_B$ \citep{sajadian2010illuminating} and $g(\Phi)$ is calculated based on the orbital motion of the planet. Thus, the net flux of the planet ($F_{\rm{p}}$) can be computed as the sum of the thermal flux ($F_{\rm{th}}$) and reflection flux ($F_{\rm{ref}}$). Eventually, we can calculate $z = {F_{\rm{p}}}/{F_{\star}}$. We note that the reflection flux is mainly important in the optical bands while the thermal flux dominates in the infra-red and sub-millimeter wavelengths. For the Jovian planets, the internal radiation of the planet due to contraction might be important. However, based on \citet{aitken19728}, the planetary thermal emission of Jupiter has pick in the range of $8-13~\mu m$, (see Fig.2 in \citet{aitken19728}). Thus, it is almost beyond the WFISRT range.\\
Without loss of generality, one can assume a circular orbit for planets. Therefore, for parameterizations of an orbit one would need the radius of orbit ($a_{\rm{p}}$), the period of orbital motion ($T$), the inclination angle between the orbital surface and the lens surface ($I$), as well as the angle between the orbit and relative velocity of source and lens star ($\beta$).\\
Using the aforementioned parameters, one can calculate the magnification by taking into account the finite size effect for both the planet and the source star. For each event, we simulate data by assuming every $15$ minutes cadence to be in the interval of $[-2t_{\rm{E}}, 2t_{\rm{E}}]$. We calculate the magnification as a function of time for both the source star and its planet.
\begin{figure*}
        \centering
        \includegraphics[width=0.49\textwidth]{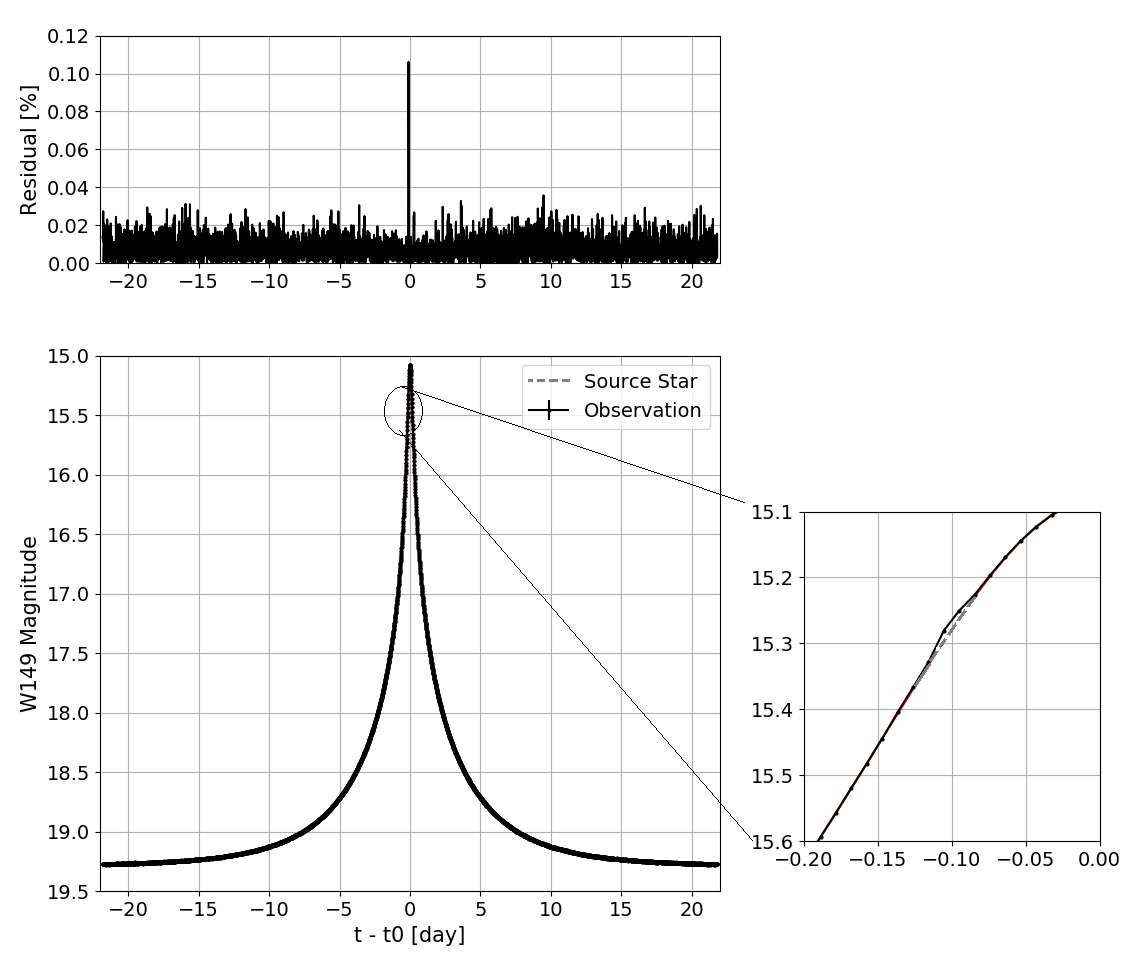}
        \includegraphics[width=0.49\textwidth]{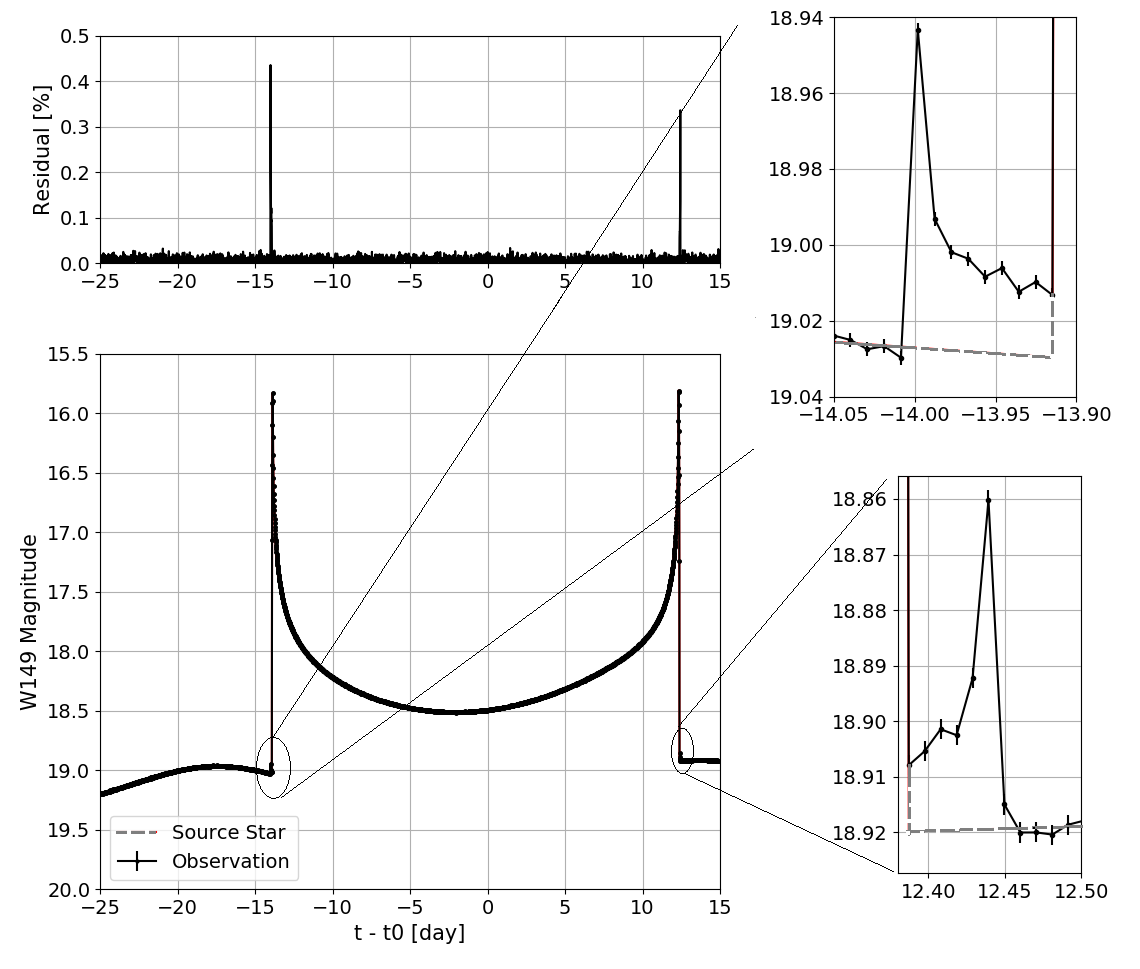}
        \caption{ The two samples of predicted final light curve in the Monte Carlo simulations of WFISRT; {\bf Left:} in single-lens case ($\rho_{\star} = 0.009$, $u_\circ = 0.02$): the predicted characteristic of the exoplanet are: orbital distance:$0.011 AU$, the mass: $968.3 M_{\oplus}$ and the temperature: $1779.6 K$ with $\Delta \chi ^{2} = 1639$, {\bf Right:} in binary-lens events ($\rho_{\star} = 0.001$, $u_\circ = 0.04$, $q = 0.85$ and $s = 1.5 R_E$): the predicted characteristic of the exoplanet are: orbital distance:$0.01 AU$, the mass: $787.9 M_{\oplus}$ and the temperature: $1863.8 K$ with $\Delta \chi ^{2} = 526$}.
        \label{fig:sample}
\end{figure*}
\begin{figure*}
        \centering
        \includegraphics[width=0.49\textwidth]{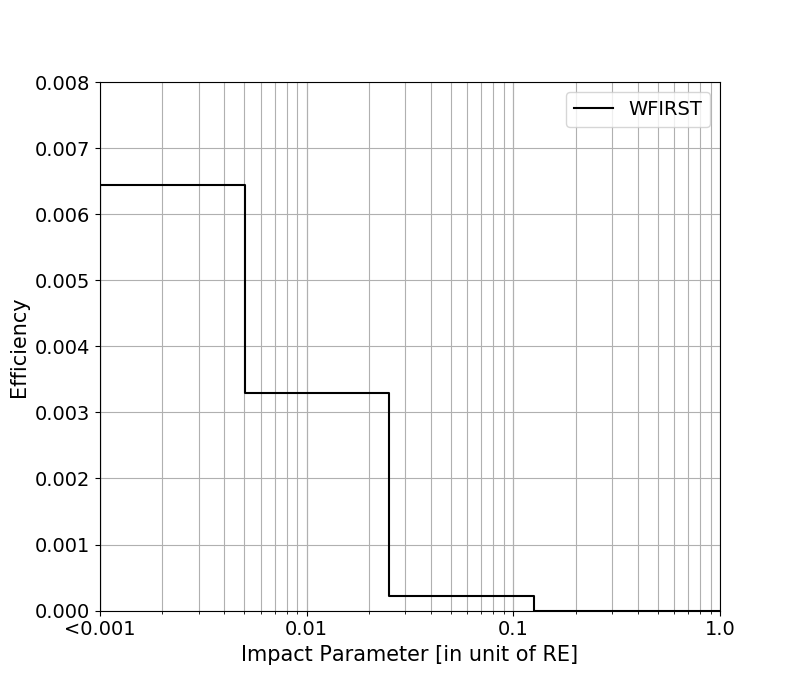}
        \includegraphics[width=0.49\textwidth]{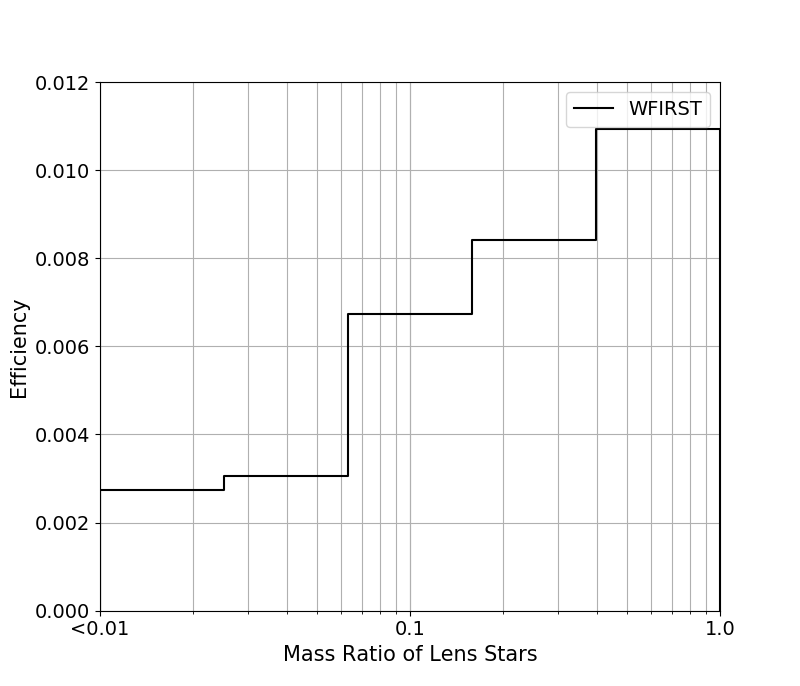}
        \caption{ The predicted distribution of the characteristics of the detectable exoplanets in the Monte Carlo simulations WFIRST. {\bf Left:} The efficiency of detectable exoplanets in terms of minimum impact parameter in single-lens events and {\bf Right:} The efficiency of detectable exoplanets in terms of the mass ratio of lens objects in the binary-lens events.}
        \label{fig:all}
\end{figure*}
\begin{figure}
        \centering
        \includegraphics[width=0.49\textwidth]{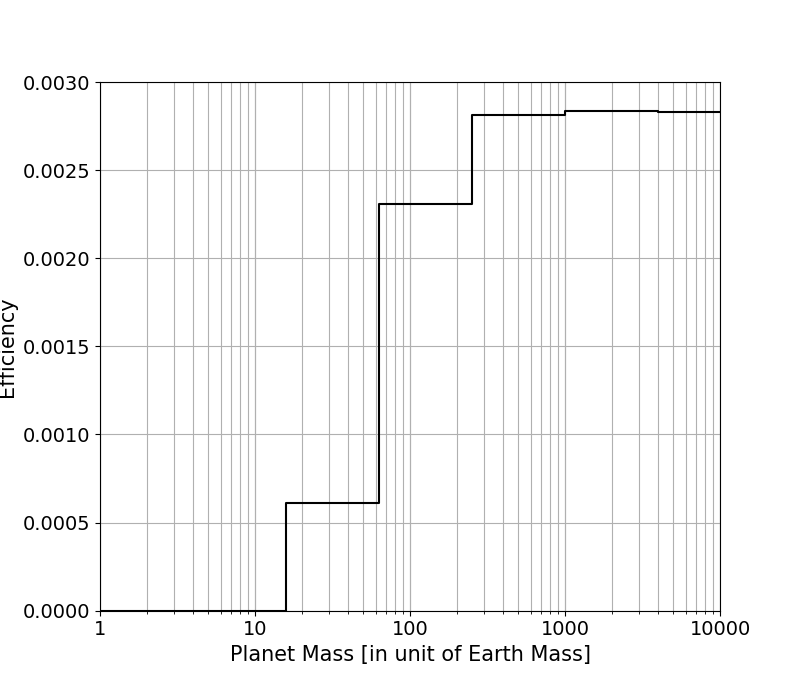}
        \includegraphics[width=0.49\textwidth]{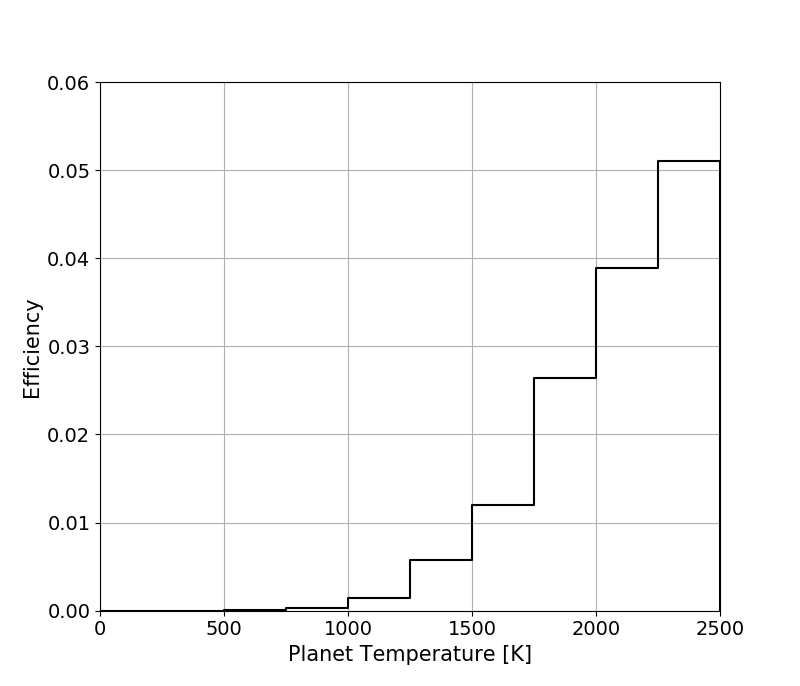}
        \includegraphics[width=0.49\textwidth]{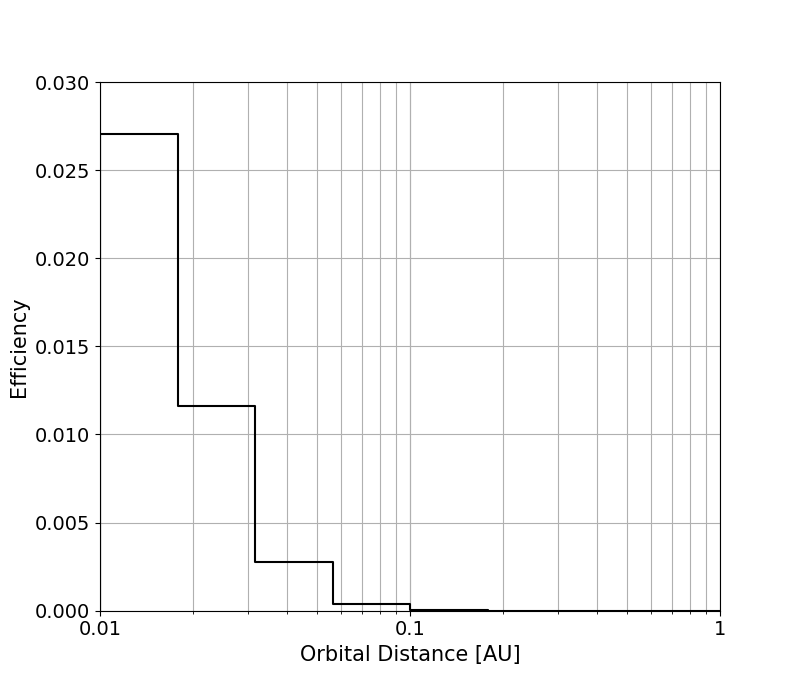}
        \caption{The predicted distributions of the mass, temperature and orbital distance of the detectable exoplanets in the Monte Carlo simulations of WFIRST from top to bottom in all simulated events. }
        \label{fig:Mass}
\end{figure}

\subsection{Detectability of the simulated events}
\label{Detectability}
The detectability of an exoplanet depends on the choice of one of the two possible models for generating light curves. In the first model, a light curve can be generated only by considering a source star with parameters $(u_{0}, t_{0}, t_{E}, \rho_{\star})$ for single-lens and $(u_{0}, t_{0}, t_{E}, \rho_{\star}, s, q)$ in case of binary-lens events, where $q$ is the mass ratio of lens objects and $s$ is the projected distance between them normalized to the Einstein radius. In both models, a light curve can be generated by including a planet in addition to the source star, it means that the models have extra parameters characterized by $(a_p, \omega_p, M_p, z, \beta, u_{\circ p}, I)$, where $a_p$, $\omega_p$, $M_p$, $\beta$, $u_{\circ p}$ are orbital distance, angular velocity, mass of planet, initial phase and impact parameter of the exoplanet, $z$ is the flux ratio of exoplanet and source star and $I$ is the inclination of the orbital plane.\\
For both models, we use $\chi^{2}$ criterion to test the detectability of a planetary signal. This criterion is based on the difference between $\chi^{2}$ of two models; We accept the hypothesis of an exoplanet detection whenever $\Delta \chi^2 \geq 150$ which is a reasonable threshold for space-based telescopes \citep{2002ApJ...574..985B,2019ApJS..241....3P,2003SPIE.4854..141B,2013MNRAS.434....2P,2014ApJ...794...52H,2012ApJ...755..102Y,2013ApJ...769...77Y}.

\section{Results}
\label{sec:results}
The results of our simulations show that the probability of detecting rocky planets by both configurations, single-lens and binary-lens, is negligibly small, virtually zero by WFIRST. Therefore, \jlike planets have the highest probabilities of detection under both configurations. This is mainly due to their higher temperatures compared to rocky planets, which leads to a significant amount of their radiation to be in or near the infrared.\\
In Figure \ref{fig:sample}, we show two samples of microlensing events, single (left panel) and binary-lens (right-panel), in which the source stars are planetary systems. In each panel, the light curve of the source star itself is shown with dashed curve and the overall light curve with the solid curve. In both light curves, the planetary signals are detectable. As seen in the light curves, the planetary signals are very small and can be misinterpreted as parallax, or orbital-motion of binary sources \citep{2009MNRAS.392.1193R} so-called Xallarap, or triple-lensing and possible other perturbation effects.  It is worthful to find all degenerate models and to estimate the number of events with such small perturbations in WFIRST data. However, our aim here is to estimate the statistics of the exoplanet signals from this microlensing channel. Once the observation with WFIRST is done, the light curves should be compared with all possible models.

Figure (\ref{fig:all}-left) illustrates the rate of the detected exoplanets as a function of the minimum impact parameter seen from WFIRST in single lens events.
A smaller impact parameter implies a higher number of detected exoplanets.
Right panel of Figure (\ref{fig:all}) shows the detection rate of exoplanets in terms of the mass ratio of lens stars in binary lens events. Again, as expected the rate is higher when the mass ratio of lens stars is closer to $1$. In this case, the caustic lines from the binary lens produce the largest area on the source plane \citep{1999A&A...349..108D}.\\
Our Monte Carlo simulations of WFIRST data predict, under the assumption of single lens configuration, that the number of detecting a \jlike planet in the distance range of $0.01$ to $10 A.U.$ of the parent star, is about $\approx 0.012\%$ with confidence level $95\%$. For the binary-lens simulation this number increases to $\approx 0.9\%$ with confidence level of $95\%$.\\
In Figure (\ref{fig:Mass}) we plot the efficiency function for detection of events in terms of the mass, temperature and orbital distance of the planets orbiting the source stars for Single-lens (dotted line) and binary-lens (solid line) channels from top to bottom, respectively.
Accordingly, (i) the heavier planets have a larger size and as a result, have higher radiation,
(ii) the hotter exoplanets have higher thermal radiation and  (iii) the closer exoplanets to their host stars are hotter as well as have more reflected flux.

\section{Discussion and Concluding Remarks}
\label{sec:discussion}
One of the merits of microlensing as a planet-finding technique is that it is the only technique that can routinely detect the exoplanets which are far from us and located in the center of Milky Way galaxy. In the conventional method, the planet orbiting around the lens star, as a binary lens produces caustic features and caustic crossing of the source star has the detection signature of the planets. Also, in the binary-lens system, detection of the exoplanets is independent of the type of host-star. Thus, this is a very unique tool to detect exoplanets which could not be hunted by the other methods. In addition, the microlensing technique is most sensitive to planets in the region of $\approx 1-10$ AU over a wide range of masses where other planet detection techniques lack the sensitivity to detect them.
 However, the $1-10$ AU region is perhaps the most important region of proto-planetary disks and planetary systems for determining their formation and subsequent evolution, and can be the habitable domain for the lens-star.

In this paper we consider another channel for detection of exoplanets where the exoplanet is orbiting around the source star. Here we took, the two major configurations in microlensing events: single and binary-lens systems in which the source star has an exoplanet. However, there are some possibility to have more than two lens objects \citep{Dank2019} and also, some non-negligible possibility to have the multi-planetary systems such as solar system. We only assume the source star has an exoplanet and do not consider the possibility of the detection of exoplanets when it contributes as one of the lens objects in the binary lens system. So far about $90$ exoplanets were detected via microlensing events \footnote{\url{https://exoplanet.eu}}. The exoplanets were mostly detected due to the caustic crossing in binary-lens events in which the exoplanet is one of the lens objects.\\
We have performed a detailed simulation of WFIRST observations in order to estimate the planet detection yield of its microlensing survey in our new observational channel. Having done so, we estimate that the probability of the exoplanet detection is $\approx 0.00012$ in single lens and $\approx 0.009$ in binary lens microlensing events in WFIRST observation. One should note that the probability of exoplanet detection with this method is a conditional probability and it depends on the probability of detection of the microlensing event
\begin{eqnarray}\label{probability}
P &=& P(\text{exoplanet detection}|\text{microlensing detection}) \\
&=& P_{1}(\text{microlensing detection}) \times P_{p}(\text{exoplanet detection}).\nonumber
\end{eqnarray}
Hence, the number of exoplanets that will be detected via WFIRST satellite through this channel of microlensing observation can be given by:
\begin{eqnarray}
N_{p, sat}=T_{obs} ~N_{\star}~\Gamma_{sat}
\end{eqnarray}
where $T_{obs}$ is the observational time,  $N_{\star}$ is the number of background stars detectable by WFISR and the $\Gamma_{sat}$  is the rate of microlensing events with planetary signal detectable by the satellite which can be given by:
\begin{eqnarray}
\Gamma_{sat }= \frac{2}{\pi} <\frac{\epsilon_{WFIRST}(t_{E})}{t_{E}}> \tau~P_{p}~P_{d}
\end{eqnarray}
where, $\epsilon_{WFIRST} (t_{E})$ is the WFIRST efficiency for detecting microlensing events with the duration of $t_{E}$, and $P_{d}$ is the probability of deflector being binary lens ($P_b$) or single lens ($P_s$) and $P_b = 0.3$ and $P_s = 0.7$ assuming that $30\%$ of microlensing events are binary and $70\%$ are single lens.
\begin{table}
    \begin{tabular}{llrrrr} \toprule
     \parbox[t]{5cm}{\textbf{WFIRST Microlensing Survey for Binary-Source Events}} \\ \midrule
    Stars ($W149 < 25$)                             & $240 \times 10^{6}$          \\
    Optical Depth                                   & $2.4 \times 10^{-6}$      \\
    Microlensing Event ($|u_\circ| <= 1$)           & $\approx 27000$              \\
    Planet Detection                              & $\approx 75$
    \end{tabular}
    \caption{The first line represents the expected total number of microlensing events for six seasons where
    each season has $72$ days monitoring \citep{2019ApJS..241....3P} . The second line is the optical depth
    toward $b = -1.5$ and $l = 0.5$ . The third line is the expected number of overall microlensing events
    and the fourth line is the number of exoplanets that can be detected via planet-source channel.}
    \label{tableOFresults}
\end{table}

We have performed the simulation toward $l=0.5 ^{\circ}$ and $b=-1.5^{\circ}$. In this direction, according to \citep{2019ApJS..241....3P} the number of source stars detectable by WFIRST with magnitude bellow than 25 is $240\times 10^{6}$. The optical depth is $\tau = 2.4\times 10^{-6}$  averaged over the distance of source stars  \citep{2019arXiv190602210M}. Form our simulation $<\epsilon_{WFIRST}/t_{E}>= 0.035$.
Hence, WFIRST satellite will detect around $\approx 3$  exoplanets (from binary lensing) and $\approx 70$ exoplanets (from single lensing) toward the Galactic bulge.  We summarize the results in Table (\ref{tableOFresults}). The number of exoplanets for different ranges of mass, temperature, and orbital distance for the single lens and binary lens events are also given in Table (\ref{table2}).
These exoplanets that could be detected are Jovian-like planets and located in the galactic bulge, which can provide an understanding of the planet formation in this part of Galaxy.

\section*{Acknowledgements}
We would like to thank the referee for their useful comments. We also thank Matthew Penny for kindly providing us the detection efficiency of WFIRST for microlensing observations. This research was supported by Sharif University of Technology Office of Vice President for Research under Grant No. G950214.

\begin{table*}
    \begin{tabular} {lllllrrrrrrrrr}\toprule 
     {\textbf{Estimated Number of Detectable Exoplanets}}                                                                      \\ \midrule
    \textbf{Mass}         & $\bf{< 100~ M_\oplus}$   & $\bf{100- 1000~ M_\oplus}$   & $\bf{1000-3000~ M_\oplus}$         & $\bf{3000-10^{4}~ M_\oplus}$ \\\midrule
    Planet Detection        & $0$                      & $0.22$                       & $7.01$                             & $68.01$   \\          \midrule
    \textbf{Temperature}  &$\bf{< 1000}$ \textbf{K}  &$\bf{1500- 2000}$ \textbf{K}  & $\bf{1500- 2000}$ \textbf{K}       & $\bf{2000- 2500}$ \textbf{K}    \\\midrule
    Planet Detection        & $0.0$                   & $0.93$                       & $9.02$                             & $65.29$\\ \midrule
    \textbf{Orbit}        & $\bf{ < 0.05}$  \textbf{A.U.}  & $ \bf{0.05 - 0.1}$  \textbf{A.U.}  & $\bf{0.1- 0.5}$  \textbf{A.U.}              & $\bf{0.5 - 1.}$  \textbf{A.U.} \\ \midrule
    Planet Detection        & $75.22$                  & $0.02$                       & $0.0$                             & $0.0$
    \end{tabular}
    \caption{The predicted numbers of detectable exoplanets in different ranges of the mass, temperature and orbital distance based on our Monte Carlo simulations of WFIRST. The total number of detections is given in Table \ref{tableOFresults}.}
   \label{table2}
   \end{table*}

\bibliographystyle{mnras}
\bibliography{main}

\end{document}